\documentclass{ws-procs9x6}

\begin{document}

\title{THE SURPRISING CONNECTION BETWEEN\\
EXACTLY SOLVED LATTICE MODELS AND\\
DISCRETE HOLOMORPHICITY}

\author{MURRAY T. BATCHELOR}

\address{Mathematical Sciences Institute, and\\
Department of Theoretical Physics, Research School of Physics and Engineering,\\
The Australian National University,\\
Canberra, ACT 0200, Australia\\
E-mail: Murray.Batchelor@anu.edu.au}

\begin{abstract}
Over the past few years it has been discovered that an ``observable'' can be set up on the lattice 
which obeys the discrete Cauchy-Riemann equations. The ensuing condition of discrete holomorphicity 
leads to a system of linear equations which can be solved to yield the Boltzmann weights of the 
underlying lattice model. Surprisingly, these are the well known Boltzmann weights which satisfy 
the star-triangle or Yang-Baxter equations at criticality. This connection has been observed for a 
number of exactly solved models. I briefly review these developments and discuss 
how this connection can be made explicit in the context of the Z$_N$ model. I also discuss how 
discrete holomorphicity has been used in recent breakthroughs in the rigorous proof of some key results 
in the theory of planar self-avoiding walks. 
\end{abstract}


\bodymatter

\section{Introduction}

The study of exactly solved lattice models in statistical mechanics has been a richly rewarding area of activity over many decades. 
Among the pantheon of exactly solved models is the two-dimensional Ising model, which is fundamental to the theory 
of phase transitions and critical phenomena \cite{baxter82,mccoy10,mussardo10}.
Two-dimensional lattice models with local interactions like the Ising model possess remarkable properties at criticality. 
The fluctuations in such systems are conformally invariant at the critical point. 
One of the triumphs of statistical physics is that this property entails the 
classification of universality classes of second order phase transitions 
in terms of conformal field theories \cite{mussardo10}. 
Roughly speaking, there is an exactly solved model for each universality class. 
The beauty of exactly solved models is that their critical Boltzmann weights satisfy the 
star-triangle or Yang-Baxter equation \cite{baxter82,mccoy10,mussardo10}.

A remarkable link has been uncovered between Yang-Baxter integrability and the condition of discrete holomorphicity \cite{rc07, ic09,cardy09}.
This involves the establishment of a local identity which is satisfied by an operator defined on the lattice. 
This operator is an example of a discrete parafermionic observable appearing in conformal field theory. 
The observable obeys a lattice version of the Cauchy-Riemann equations and is thus known to be discretely holomorphic. 
The condition of discrete holomorphicity leads to a system of linear equations which can be solved to yield the 
Boltzmann weights of the lattice model. 
These are the Boltzmann weights which are known to satisfy the Yang-Baxter equations at criticality. 
This connection has been observed for a number of exactly solved models, \cite{rc07, ic09,cardy09}
including the Z$_N$ model,\cite{rc07} the Potts model \cite{ic09} and the O$(n)$ loop model \cite{ic09}.

Here we will discuss these developments in terms of the O$(n)$ loop model. 

\section{The O$(n)$ loop model}

\begin{figure}[h]
\begin{center}
\psfig{file=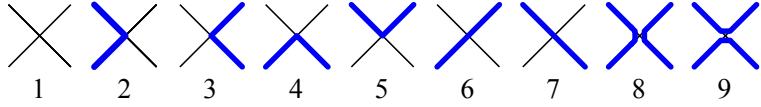,width=10cm}
\end{center}
\caption{The Boltzmann weights of the O$(n)$ loop model.}
\label{fig:weights}
\end{figure}

The O$(n)$ loop model on the square lattice was originally constructed by Nienhuis \cite{nienhuis90,bn89}.
It is exactly solved by means of the Bethe Ansatz, with periodic \cite{bnw89} and open \cite{yb95, mixed} boundaries. 
As an integrable model, its underlying $R$-matrix is that of the Izergin-Korepin, \cite{ik} 
or $A_2^{(2)}$ vertex model \cite{yb95}.
The partition function is given by 
\begin{equation}
Z_{\rm loop} = \sum_{\mathcal G} \rho_1^{m_1} \ldots \rho_9^{m_9} \, n^P 
\end{equation}
which is a sum over all possible loop configurations $\mathcal G$. 
Here $m_i$ is the number of occurrences of Boltzmann weight $\rho_i$ 
and $P$ is the total number of closed loops of fugacity $n$ in any given configuration. 
The nine non-zero Boltzmann weights $\rho_i$ are shown in Fig.~\ref{fig:weights}. 
These weights are given by \cite{nienhuis90,yb95}
\begin{align}
\rho_1 &= \sin (3\lambda -u) \sin u + \sin 2\lambda \sin 3\lambda \nonumber\\
\rho_2 &= \rho_3 = \epsilon_1 \sin(3\lambda -u) \sin 2\lambda \nonumber\\
\rho_4 &= \rho_5 = \epsilon_2 \sin u \sin 2\lambda \nonumber\\
\rho_6 &= \rho_7 = \sin(3\lambda -u) \sin u \label{weights}\\
\rho_8 &= \sin(3\lambda -u) \sin(2\lambda -u) \nonumber\\
\rho_9 &= - \sin(\lambda -u) \sin u \nonumber
\end{align}
with $\epsilon_1^2 = \epsilon_2^2 =1$, $0 < u < 3\lambda$ and $n = - 2 \cos 4\lambda$. 

The parafermion operator is defined on the lattice as \cite{ic09} 
\begin{equation}
F(z) = \sum_{\mathcal G \in \Gamma(0,z)} P({\mathcal G}) {\rm e}^{-{\rm i} \sigma W(z)} .
\end{equation}
Here $P(\mathcal G)$ is the probability of configuration $\mathcal G$, $\Gamma(0,z)$ is a set of loop 
configurations for which points $0$ and $z$ belong to the same loop, $W(z)$ is the winding angle of this loop 
from points $0$ and $z$, and $\sigma$ is the spin of the parafermion $F(z)$. 
For the O$(n)$ loop model 
\begin{equation}
F(z) = \sum_{\gamma(a \to z) \in \Omega} {\rm e}^{-{\rm i} \sigma W(\gamma(a\to z))} \rho_1^{m_1} \ldots \rho_9^{m_9} \, n^P 
\end{equation}
for some path $\gamma(a \to z)$ across a domain $\Omega$. 

\begin{figure}[h]
\begin{center}
\psfig{file=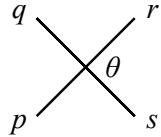,width=2cm}
\end{center}
\caption{The labelling of a vertex.}
\label{fig:vertex}
\end{figure}

What Ikhlef and Cardy \cite{ic09} have shown is that the parafermion $F(z)$ is discretely holomorphic, i.e., it satisfies 
\begin{equation}
F(p) - {\rm e}^{{\rm i} \theta} F(q) - F(r) +  {\rm e}^{{\rm i} \theta} F(s) = 0
\end{equation}
on each vertex as defined in Fig.~\ref{fig:vertex}. 
This condition is equivalent to requiring that $F$ is divergence and curl free. 

\begin{figure}[h]
\begin{center}
\psfig{file=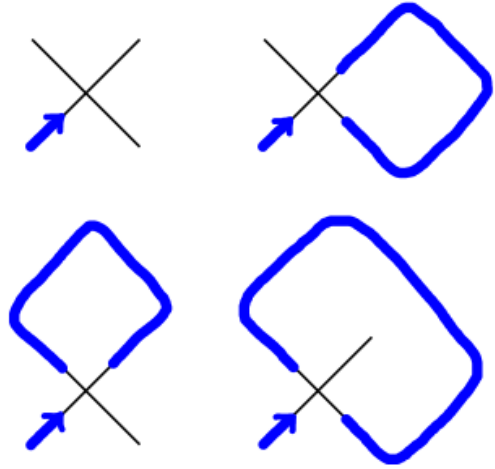,width=7cm}
\end{center}
\caption{The four possible external connectivities from mid edge $p$.}
\label{fig:loops}
\end{figure}

To establish this, one chooses a particular mid edge $p$. 
There are four possible different external connectivities, as shown in Fig.~\ref{fig:loops}.
These possibilities result in four equations 
\begin{align}
\rho_1 + \nu \rho_2 - \nu \zeta^{-1}  \rho_4 - \rho_7 &=0 \nonumber\\
- \zeta^{-1} \rho_2 + n \rho_5 + \nu \zeta \rho_7 - \nu \zeta^{-1} (\rho_8 + n \, \rho_9) &=0 \nonumber\\
n\, \rho_3 - \zeta \rho_4 - \nu \zeta^{-2}  \rho_7 + \nu (n\, \rho_8 + \rho_9) &=0 \label{relations}\\
- \nu  \zeta^{-2} \rho_2 + \nu \zeta \rho_4 + n\, \rho_6 - \zeta^{-2}  \rho_8 - \zeta^2 \rho_9 &=0 \nonumber
 \end{align}
where $\nu={\rm e}^{{\rm i} \theta(\sigma+1)}$ and $\zeta={\rm e}^{{\rm i} \sigma \pi}$.
The remarkable point is that Ikhlef and Cardy \cite{ic09} were able to solve this set of linear equations to yield the  
Boltzmann weights $\rho_1, \ldots, \rho_9$ and the conformal spin 
\begin{equation}
\sigma = 1 - \frac{3\lambda}{\pi} \label{sigma}
\end{equation}
provided $u = \sigma(\theta-\pi) + \theta$. 
In particular, at $\theta=\frac{\pi}{2}$, $u=3\lambda/2$, the isotropic point. 
Substitution of the Boltzmann weights (\ref{weights}) into equations (\ref{relations})  
shows that these equations are indeed satisfied.

The surprising point is that the Yang-Baxter equations are {\em cubic} in the Boltzmann weights, 
whereas imposing discrete holomorphicity led to a set of equations which are {\em linear} in the Boltzmann weights.
There thus appears to be a simpler route to uncovering integrable lattice models at criticality.
Why is this? And what is the precise link between discrete holomorphicity and integrability? 
These questions will be taken up further in Section 4 in the context of the Z$_N$ model, 
which can be discussed from an algebraic point of view.

We now turn to the related developments in the theory of self-avoiding random walks.

\section{Self-avoiding random walks}

The random walk, also known as the drunkard's walk, was introduced over 100 years ago \cite{pearson}. 
The study of random walks has since developed into the central pillar of the theory of probability and random processes \cite{hughes}. 
A key feature of the random walk is that, like the drunkard, it has no memory of where it has been. 
This absence of memory is the defining property of a Markovian process.

A more challenging class of mathematical problems arise when a random walk is not allowed to revisit a site it has already visited. 
Such self-avoiding random walks were originally proposed as models of long polymer chains in a good solvent. 
Despite the simplicity of their definition, the non-Markovian nature of self-avoiding random walks makes them notoriously difficult to study. 
Nevertheless much is known about the theory of self-avoiding walks, particularly in higher dimensions \cite{ms}. 
Much is also known in two and three dimensions from extensive computer enumeration. 
The self-avoiding constraint is more restrictive in two dimensions and thus self-avoiding walks on planar lattices are arguably more interesting to study. 

\begin{figure}[h]
\begin{center}
\psfig{file=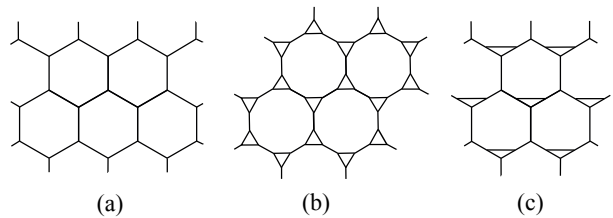,width=8cm}
\end{center}
\caption{The (a) honeycomb, (b) 3-12 and (c) martini lattices.}
\label{fig:lattices}
\end{figure}

In 1982 Nienhuis \cite{nienhuis82} was able to obtain some key information for self-avoiding walks 
by considering the O$(n)$ loop model on the honeycomb (hexagonal) lattice depicted in Fig.~\ref{fig:lattices}(a). 
In general, the  O$(n)$ model has a connection to self-avoiding walks in the limit $n = 0$ \cite{degennes}. 
The critical Boltzmann weights of the model on the honeycomb lattice were obtained from an exact mapping to the Potts model \cite{nienhuis82}. 
The Boltzmann weights defining the critical line of the O$(n)$ loop model on the honeycomb lattice also follow as a condition for the 
model to be solved exactly in terms of the Bethe Ansatz \cite{baxter86,bb88, suzuki}. 
More generally the O$(n)$ loop model on the honeycomb lattice follows as a limiting case of the exactly solved O$(n)$ loop model 
on the square lattice.\footnote{Specifically, when either $u=\lambda$ or $u=2\lambda$ in (\ref{weights}), corresponding to the two orientations 
in Fig.~\ref{fig:orient}.}

The critical Boltzmann weights of the model give the value for the connective constant $\mu$, 
which is a measure of the growth in the number of allowed self-avoiding walks with increasing length. 
The Nienhuis result, 
\begin{equation}
\mu = \sqrt{2+\sqrt 2} = 1.847~759\ldots
\label{mu}
\end{equation}
was generally believed to be exact, having been confirmed to high precision from extensive enumeration of self-avoiding walks on the honeycomb lattice. 
No such exact results are known for the other regular planar lattices.

Similar exact results are known for self-avoiding walks on related semi-regular lattices. 
These are the 3-12 lattice (Fig.~\ref{fig:lattices}(b)) and the martini lattice  (Fig.~\ref{fig:lattices}(c)). 
For the 3-12 lattice, the exact value \cite{jg98,batch98} $\mu = 1.711~041\ldots$ follows from the solution of  
\begin{equation}
\frac{1}{\mu^2} + \frac{1}{\mu^3} = \frac{1}{\sqrt{2+\sqrt 2}} .
\end{equation}
And for the martini lattice, the exact value \cite{dfg12} $\mu =1.750~564\ldots$ follows from the solution of 
\begin{equation}
\frac{1}{\mu^3} + \frac{1}{\mu^4} = \frac{1}{\sqrt{2+\sqrt 2}} .
\end{equation}
More generally, the O$(n)$ model on the honeycomb lattice can be mapped to the O$(n)$ model on 
the 3-12 \cite{batch98} and martini lattices \cite{dfg12}.

Up until recently, a rigorous mathematical proof of the result (\ref{mu}) had seemed out of reach. 
In a remarkable paper, Duminil-Copin and Smirnov \cite{ds12} have given an ingenious proof of the Nienhuis result. 
The first part of the proof involves defining a parafermionic observable on the honeycomb lattice 
which is shown to be discretely holomorphic. 
The Boltzmann weights of the model follow from the condition of  discretely holomorphicity. 
In this sense the value (\ref{mu}) also appears as a consequence of discrete holomorphicity. 
The second part of the proof builds on previously hard won results in the general theory of self-avoiding 
walks to rigorously establish that this value is indeed the connective constant.

\subsection{The polymer adsorption transition}

\begin{figure}[h]
\begin{center}
\psfig{file=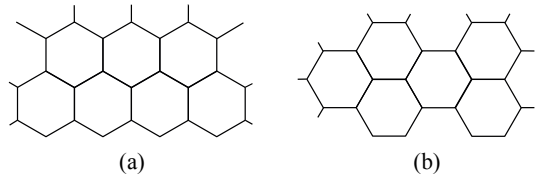,width=7cm}
\end{center}
\caption{The two orientations of the honeycomb lattice.}
\label{fig:orient}
\end{figure}

These arguments have been extended \cite{beaton12a,beaton12d} to provide a rigorous proof of the critical surface fugacity $y_c$ 
of self-avoiding walks on the honeycomb lattice in the presence of a boundary. 
This is the critical adsorption temperature at which the polymer becomes attached to the surface. 
For $y>y_c$ the polymer is adsorbed and for $y<y_c$ the polymer is desorbed.
The critical value depends on the orientation of the lattice (see Fig.~\ref{fig:orient}).

For the orientation shown in Fig.~2(a), the exact value \cite{by95}
\begin{equation}
y_c = 1 + \sqrt 2 = 2.414\ldots
\label{kappa1}
\end{equation}
had been obtained earlier using the underlying integrability of the O$(n)$ loop model with a boundary \cite{yb95}. 
For the orientation shown in Fig.~2(b), the exact value \cite{bbwo98} 
\begin{equation}
y_c = \sqrt{\frac{2+\sqrt 2}{1+\sqrt 2 - \sqrt{2+\sqrt 2}}} = 2.455\ldots 
\label{kappa2}
\end{equation}
also followed from the integrable boundary weights of the O$(n)$ loop model \cite{yb95}. 

Building on the approach of Duminil-Copin and Smirnov, these
results have been shown by Beaton {\it et al.} \cite{beaton12a,beaton12d} to also follow from the discrete holomorphicity of a 
parafermionic observable. 
Their mathematical proof that these values are indeed the critical adsorption temperatures is an impressive {\em tour de force}, 
involving the adaption of previous rigorous work on self-avoiding walks and the rigorous treatment of 
self-avoiding bridges in a finite strip.

We turn now to further examine the connection between discrete holomorphicity and integrability.

\section{The Z$_N$ model}

The Z$_N$ model is an $N$-state model in which the spins $s_r$ at vertex $r$ take values 
from the $N$th roots of unity, $s_r = \omega^{q_r}$, with $q_r \in \{0,1,\ldots,N-1\}$ and $\omega=\exp(2\pi {\rm i}/N)$. 
The model contains a number of well known models as special cases, namely the Ising model ($N=2$), the three-state Potts model ($N=3$) 
and the Ashkin-Teller model ($N=4$).
The Z$_N$ model has a well known set of critical points obtained by Fateev and Zamolodchikov (FZ) as solutions 
to the star-triangle relations \cite{fz}. 
By defining lattice parafermions as products of neighbouring order and disorder variables with a suitable phase factor, 
it was shown by Rajapour and Cardy that the FZ Boltzmann weights ensure that the lattice parafermions obey the discrete version of the 
Cauchy-Riemann equations, i.e., they enjoy the property of discrete holomorphicity \cite{rc07,cardy09}. 
These results were extended to lattices with anisotropic interactions. 

The underlying motivation for this work on the Z$_N$ model is the construction of discretely holomorphic parafermions, which are 
expected to become the holomorphic parafermions of the conformal theory. 
Here our interest is in the connection with integrability.
From that perspective, the key finding is that the contour sum around each elementary face or rhombi of the lattice 
vanishes, the key ingredient for which is the set of FZ critical Boltzmann weights. 
This approach was extended  by explicitly considering the condition of 
discrete holomorphicity on two and three adjacent rhombi \cite{ab12}.
For two rhombi this leads to a quadratic equation for the Boltzmann weights which implies the known inversion relations 
for the Z$_N$ model. 
For three rhombi it leads to a cubic equation from which the star-triangle relation follows. 
The simplicity of the discrete holomorphic approach is that the two- and the three-rhombus equations are equivalent 
to the one-rhombus equation. 

\section{Discussion}

As alluded to in the previous section, the original motivation behind the recent progress on various aspects  of 
discrete holomorphicity covered in this article was to set up the necessary discrete building blocks for the passage to the 
continuous scaling limit. 
There is ample motivation for doing this. 
For example, the geometry of the curves traced out by self-avoiding walks is conjectured to be described by the 
mathematical theory of Schram-Loewner evolution (SLE$_{8/3}$) \cite{lsw04}.
Basically, if the scaling limit of two-dimensional self-avoiding walks exists and has a certain conformal invariance property, 
then the scaling limit must be described by the particular stochastic process SLE$_{8/3}$. 
However, establishing the existence of the scaling limit (and therefore also its conformal invariance) is a difficult open problem. 
Nevertheless, significant progress in this direction has been made for other two-dimensional models, most notably 
percolation and the Ising model. \cite{stas1,stas2,ds11}.

Although the programme to establish the scaling limit of self-avoiding walks in a mathematically rigorous way is not yet complete, 
the first step, involving establishing analytic properties on the lattice via the discrete 
Cauchy-Riemann equations, has led to some remarkable developments.
These include the surprising mathematical proofs of the results (\ref{mu}), (\ref{kappa1}) and (\ref{kappa2}) 
for the connective constant and critical surface fugacity 
for self-avoiding walks on the honeycomb lattice \cite{ds12,beaton12a,beaton12d}.

There are also ongoing developments with regard to the connection with integrability, 
particularly at a boundary.
A boundary condition can also be imposed on a discretely parafermionic observable \cite{ikhlef12,degier12}.  
This approach has led to a set of new integrable boundary weights for the Z$_N$ model \cite{ikhlef12}.
Several new sets of boundary weights have also been obtained in this way 
for the O$(n)$ and $C_2^{(1)}$ loop models \cite{degier12}.

More developments are eagerly awaited.

\section*{Acknowledgments}
This work has been partly supported by the Australian Research Council.

\end{document}